\documentclass[conference]{IEEEtran}
\IEEEoverridecommandlockouts
\usepackage{cite}
\usepackage{amsmath,amssymb,amsfonts}
\usepackage{algorithmic}
\usepackage{graphicx}
\usepackage{textcomp}
\usepackage{xcolor}
\usepackage{hyperref}

\usepackage{url}
\usepackage{booktabs}       
\usepackage{multirow}
\usepackage{amssymb}
\usepackage{pifont}
\usepackage{arydshln}
\usepackage{pgfplots}
\pgfplotsset{compat=1.18} 
\usepackage{dsfont}

\def\BibTeX{{\rm B\kern-.05em{\sc i\kern-.025em b}\kern-.08em
    T\kern-.1667em\lower.7ex\hbox{E}\kern-.125emX}}

\DeclareMathOperator*{\argmax}{arg\,max}

\begin{document}

\title{COCOLA: Coherence-Oriented Contrastive Learning of Musical Audio Representations\\
\thanks{The authors were partially supported by the ERC grant no. 802554 (SPECGEO), PRIN 2020 project no. 2020TA3K9N (LEGO.AI), PRIN 2022 project no. 2022AL45R2 (EYE-FI.AI, CUP H53D2300350-0001),  PNRR MUR project no. PE0000013-FAIR, and AWS Cloud Credit for Research program.}
}

\author{
\begin{tabular}{@{}c@{}}
Ruben Ciranni$^{1*}$ \thanks{$^*$Equal contribution.}
\qquad Giorgio Mariani$^{1*}$
\qquad Michele Mancusi$^{2*}$
\qquad Emilian Postolache$^{3*}$\\
\qquad Giorgio Fabbro$^{2}$
\qquad Emanuele Rodol\`a$^{1\dagger}$ \thanks{$^\dagger$Shared last authorship.}
\qquad Luca Cosmo$^{3\dagger}$
\end{tabular}
\IEEEauthorblockA{ \\
\\
$^1$Sapienza University of Rome, Italy \quad $^2$Sony Europe B.V., Stuttgart, Germany \\
$^3$DAIS, Ca' Foscari University of Venice, Italy \\
}
}

\maketitle

\begin{abstract}
We present COCOLA (Coherence-Oriented Contrastive Learning for Audio), a contrastive learning method for musical audio representations that captures the harmonic and rhythmic coherence between samples. Our method operates at the level of the individual stems composing music tracks and can input features obtained via Harmonic-Percussive Separation (HPS). COCOLA allows an objective evaluation of generative models for music accompaniment generation, which are difficult to benchmark with established metrics. In this regard, we evaluate recent music accompaniment generation models, demonstrating the effectiveness of our proposed method. We release the model checkpoints trained on public datasets containing separate stems (MUSDB18-HQ, MoisesDB, Slakh2100, and CocoChorales).
\end{abstract}


\section{Introduction}
\label{sec:introduction}
Recently, there have been significant advances in music generation in the continuous domain \cite{ schneider2023mousai, garcia2023vampnet, evans2024fast, evans2024stable}, thanks to the impressive development of generative models \cite{song2021scorebased, ho2020denoising}.
In addition to producing high-quality tracks of increasing length \cite{evans2024fast}, these models offer precise semantic control through textual conditioning \cite{elizalde2023clap, raffel2020exploring}. However, they are limited as tools for musical composition, since they output a final mix containing all stems. To overcome this, a new range of compositional generative models for music is emerging \cite{mariani2023multi, han2023instructme, parker2024stemgen}, where \textit{(i)} the generative tasks are defined at the stem level and \textit{(ii)} their usage is iterative/interactive. The main application of these models is accompaniment generation, where the model is required to produce a \textit{coherent} stem that blends harmonically and rhythmically with multiple conditioning sources. While pioneering models \cite{grachten2020bassnet, donahue2023singsong} can generate waveform accompaniments, their output is limited to a single stem category, and as such they cannot be used iteratively (acting sequentially on the model's outputs) in a composition process.

A significant problem with this line of research is the lack of objective metrics for quantifying the coherence of the generated outputs w.r.t. the inputs. For example, \cite{mariani2023multi} proposes the sub-FAD metric as a multi-stem generalization of the FAD \cite{roblek2019fr} protocol proposed in \cite{donahue2023singsong}. However, this metric is suboptimal for assessing coherence, as it focuses on global quality instead of the level of harmony and rhythm shared by constituent stems.
\begin{figure}[t!]
 \centerline{
 \includegraphics[width=0.8\columnwidth]{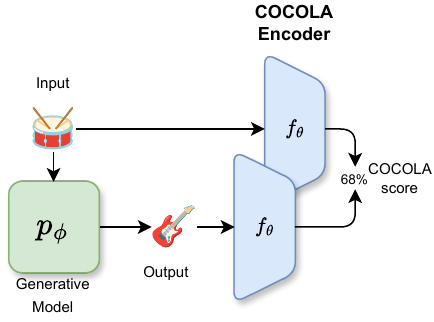}
 }
 \caption{\textbf{Illustration of COCOLA Score.} COCOLA is a contrastive model that estimates the coherence between instrumental tracks and generated accompaniments.}
 \label{fig:overview}
\end{figure}
To this end, we propose and release\footnote{\url{https://github.com/gladia-research-group/cocola}} a novel contrastive model called \textit{COCOLA (Coherence-Oriented Contrastive Learning for Audio)}, which can evaluate the coherence between conditioning tracks and generated accompaniments (Figure \ref{fig:overview}). The model is trained by maximizing the agreement between disjoint sub-components of an audio window (sub mixtures of stems) and minimizing it on sub-components belonging to different windows. With the model, we define a \textit{COCOLA Score} as the similarity between conditioning tracks and accompaniments in the embedding space, and use it to benchmark accompaniment generation models. 

After discussing related work in Section \ref{sec:related}, we introduce COCOLA in Section \ref{sec:method}. We describe the experimental setup in Section \ref{sec:setup} and present the results in Section \ref{sec:results}. We conclude the article in Section \ref{sec:conclusion}.

 \begin{figure}[!t]
 \centerline{
 \includegraphics[width=1\linewidth]
 {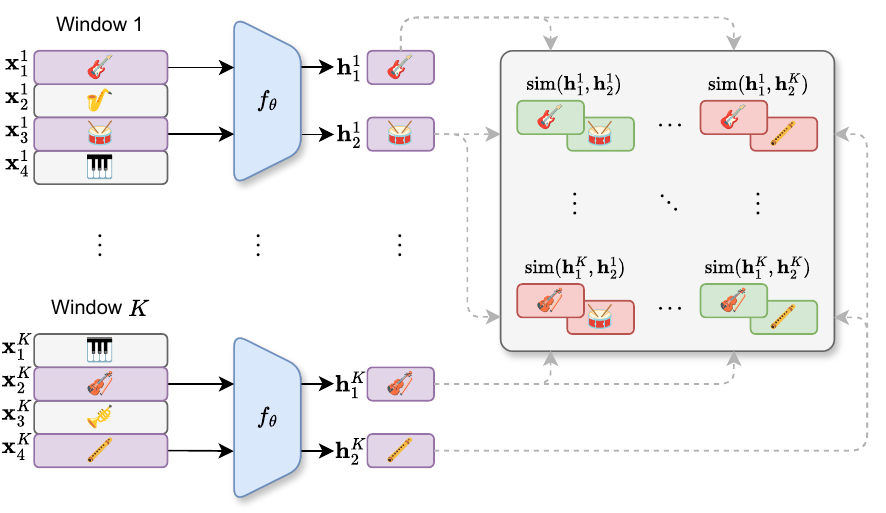}
 }
 \caption{\textbf{The COCOLA training procedure (single stem case).} 
Windows of size $L$ are randomly cropped from $K$ tracks (left). Two distinct stems per window are randomly selected (e.g., Guitar $\mathbf{x}^1_1$ and Drums $\mathbf{x}^1_3$ in the first window), embedded using the COCOLA encoder $f_\theta$, yielding latent representations (e.g., $\mathbf{h}^1_1$ and $\mathbf{h}^1_2$). Contrastive loss (Eq.~\eqref{eq:cross_entropy}) is computed with positive pairs within windows and negatives across windows. 
}
 \label{fig:cocola}
\end{figure}

\section{Related Work}
\label{sec:related}
 
\subsection{Contrastive Models for Audio}
\label{subsec:related_contrastive}
Contrastive learning \cite{chopra2005learning, oord2018representation} can be both formulated as a supervised or self-supervised problem.

Supervised contrastive learning methods often rely on cross-modal approaches, requiring labeled data alongside the audio. Early works \cite{favory2020coala} employ the contrastive loss \cite{chen2020simple} for aligning audio with simple labels. MuLaP \cite{manco2022learning} was the first model to jointly represent audio and complex text using a joint transformer encoder with cross-attention layers. Recent models \cite{manco2022contrastive,  elizalde2023clap, wu2023large} have adopted separate text and audio encoders, allowing independent use of these branches during inference. 

Self-supervised representation learning methods \cite{pascual2019learning,  huang2022masked} create embeddings from structural information in audio data. In \cite{jansen2018unsupervised}, positive examples for triplet loss are built using Gaussian noise, time/frequency translations, and time proximity, without ensuring coherence (e.g., mixing unrelated sounds). Subsequent methods \cite{spijkervet2021contrastive} adapted the contrastive loss. COLA \cite{saeed2021contrastive} uses the simple criterion of sampling positive pairs only from the same audio track, outperforming a supervised baseline. \cite{garoufis2023multi} pairs mixtures with sources extracted via source separation.

A line of research \cite{chen2020neural, huang2021modeling, lattner2022samplematch, riou2024stem} focuses on estimating the compatibility of loops for retrieval tasks. Although these approaches, like ours, aim to evaluate the coherence between audio tracks, they address a distinct task (retrieval) rather than generative model evaluation. Moreover, they have limitations: some of them do not extract individual stems \cite{chen2020neural}, while others rely on fixed stem choices \cite{huang2021modeling, lattner2022samplematch, riou2024stem}. In contrast, our method supports any combination of stems, offering greater flexibility.


\subsection{Waveform Music Accompaniment Generation Models}
MSDM \cite{mariani2023multi}, based on diffusion models \cite{ho2020denoising, song2021scorebased} was the first model to generate waveform accompaniments belonging to different stem classes. Following, state-of-the-art models \cite{parker2024stemgen, nistal2024diff} took the form of conditional generative models which respond with an output stem to an input track. Based on generative source separation via Bayesian inference \cite{jayaram2020source, zhu2022music, postolache2023latent}, GMSDI \cite{postolache2024generalized} performs the same tasks as MSDM by enabling source separation during generation, requiring models trained solely on mixtures and text.

Another category of models \cite{han2023instructme, zhang2024instruct}, following AUDIT \cite{wang2024audit}, perform music editing by modifying the input track. Such models can add instruments to a track, but given the generative nature of the model, isolating such stems via the difference with the input or source separation can introduce artifacts.

\label{subsec:related_accompaniment}

\section{Method}

\label{sec:method}
\subsection{Stem-Level Contrastive Learning}
\label{subsec:method_contrastive}
In our setting, we have access to a dataset $D = \{ \bar{\mathbf{x}}^k \}_{k=1, \dots, \bar{K}}$ containing $\bar{K}$ musical tracks $\bar{\mathbf{x}}^k$, each separated into a variable number $N$ of individual stems $\bar{\mathbf{x}}^k_n$, i.e., $\bar{\mathbf{x}}^k = \{\bar{\mathbf{x}}^k_n \}_{n=1,\dots, N}$. As a first step, we sample a batch of $K < \bar{K}$ tracks $\{ \bar{\mathbf{x}}^k \}_{k=1, \dots, K}$ from $D$, with possible repetitions. Following, we slice a window $\mathbf{x}^k$ of size $L$ for each track $\bar{\mathbf{x}}^k$ in the batch (all stems in a window share the same length), such that no window contained in the same track overlaps for more than a ratio $r$, obtaining a new batch $\{\mathbf{x}^k \}_{k=1,\dots, K}$. Afterward, we select, for each $k$, two disjoint non-empty stem subsets $X_1^{k}, X_2^{k}$ of $\mathbf{x}^k$. We define the submixes $\mathbf{m}^k_1$ and $\mathbf{m}^k_2$ as the sum of the stems in $X_1^{k}, X_2^{k}$:
\begin{equation}
\label{eq:sub_mixes_contrastive}
    \mathbf{m}^k_1 = \sum_{\mathbf{x}_n^k \in X_{1}^k} \mathbf{x}^k_n, \qquad \mathbf{m}^k_2 = \sum_{\mathbf{x}_n^k \in X_{2}^k} \mathbf{x}^k_n\,.
\end{equation}
When $X_1^{k}, X_2^{k}$ are singletons, the submixes are simply two stems in the window (single stem case). 

Like in COLA\cite{saeed2021contrastive}, we use a convolutional audio-only encoder\footnote{In our notation, we incorporate into $f_\theta$ any domain transform preceding or following the convolutional network operations, like the (pre) mel-filterbank map and the (post) projection head $g$ in COLA.} $f_\theta: \mathbb{R}^L \to \mathbb{R}^d$, mapping $\mathbf{m}^k_1$ and $\mathbf{m}^k_2$ to lower-dimensional embedding vectors $\mathbf{h}^k_1 = f_\theta(\mathbf{m}^k_1)$ and $\mathbf{h}^k_2 = f_\theta(\mathbf{m}^k_2)$, with $d$ the embedding dimension. The COCOLA training procedure maximizes the agreement between pairs $\mathbf{h}^k_1, \mathbf{h}^k_2$ of embeddings belonging to the same window while minimizing it for pairs $\mathbf{h}^k_1, \mathbf{h}^j_2$ ($j \neq k$) belonging to different windows. As in  COLA, we use a bilinear similarity metric:
\begin{equation}
\label{eq:similarity}
    \text{sim}(\mathbf{h}^k_1,\mathbf{h}^j_2) = (\mathbf{h}^k_1)^T \mathbf{W} \mathbf{h}^j_2 \,,
\end{equation}
where $\mathbf{W}$ is a learnable matrix. 
The loss we optimize is the multi-class cross entropy:
\begin{equation}
\label{eq:cross_entropy}
\mathcal{L} = - \sum^{K}_{k=1}\log \frac{\exp(\text{sim}(\mathbf{h}^k_1, \mathbf{h}^k_2))}{\sum^K_{j=1} \exp(\text{sim}(\mathbf{h}^k_1, \mathbf{h}^j_2))}\,.
\end{equation} 

\begin{table}[!t]
\centering
\caption{\label{tab:classification} Classification accuracy tests (\%) with COCOLA models using $K=2$ sub-mixture test pairs  (higher is better). MUSDB18-HQ is used as a hold-out test dataset.}
\resizebox{\columnwidth}{!}{\begin{tabular}{@{}lllll@{}}
\toprule
             & \multicolumn{4}{c}{\textbf{Test Dataset}} \\ \midrule
\textbf{Model} & \multicolumn{1}{l}{MUSDB18-HQ} & \multicolumn{1}{l}{MoisesDB} & \multicolumn{1}{l}{Slakh2100} & \multicolumn{1}{l}{CocoChorales} \\ \midrule
\textbf{w\textbackslash o HPS} & & & &  \\
MoisesDB  \cite{pereira2023moisesdb}    &    52.56   &   53.01     &    51.22     &    60.32    \\
Slakh2100  \cite{manilow2019cutting}  &     53.06    &    53.58  &    53.78  &    59.35   \\
CocoChorales \cite{wu2022chamber} &   70.10      &  61.48       &      67.50   &    99.78   \\
All          & 90.43   & 93.06  & 90.06   & \textbf{99.89} \\
\textbf{w\textbackslash \ HPS} & & & &  \\ 
All          & \textbf{93.87}   & \textbf{93.67}  & \textbf{94.27}   & 99.68
\\ \bottomrule
\end{tabular}}
\end{table}

\begin{table}[!t]
\centering
\caption{\label{tab:classification_batch} Classification accuracy tests (\%) with  COCOLA HPS All varying the values of $K$.}
\begin{tabular}{@{}lllll@{}}
\toprule & \multicolumn{4}{c}{\textbf{Test Dataset}} \\ \midrule
\textbf{K} & \multicolumn{1}{l}{MUSDB18-HQ} & \multicolumn{1}{l}{MoisesDB} & \multicolumn{1}{l}{Slakh2100} & \multicolumn{1}{l}{CocoChorales} \\ \midrule
 8  & 78.02  &  73.68  & 79.72 & 98.27 \\  
16      &    70.32    &    62.17     &  72.62 & 96.67      \\
64     &    54.33  &    34.04  &    59.35 & 90.67     \\ \bottomrule
\end{tabular}
\end{table}

\begin{table*}[t!]
\caption{Comparison between music accompaniment generation models.}
\label{tab:generative}
\centering
\begin{tabular}{@{}lllllll@{}}
\toprule
\textbf{Method} &
  \multicolumn{1}{l}{\begin{tabular}[l]{@{}l@{}}FAD $\downarrow$ \\ CLAP\end{tabular}} &
  \multicolumn{1}{l}{\begin{tabular}[l]{@{}l@{}}FAD $\downarrow$\\ EnCodec\end{tabular}} &
  \multicolumn{1}{l}{\begin{tabular}[l]{@{}l@{}}FAD $\downarrow$\\ VGGish\end{tabular}} &
  \multicolumn{1}{l}{\begin{tabular}[l]{@{}l@{}}COCOLA Score$\uparrow$ \\ Percussive + Harmonic\\ \end{tabular}} &  \multicolumn{1}{l}{\begin{tabular}[l]{@{}l@{}}COCOLA Score $\uparrow$ \\ Percussive \end{tabular}} &  \multicolumn{1}{l}{\begin{tabular}[l]{@{}l@{}}COCOLA Score $\uparrow$\\ Harmonic \end{tabular}}  \\ \midrule
\textbf{MoisesDB Test} & & & & &   \\
Random & \textbf{0.072}  & 19.37 & \textbf{0.38} & 50.53 & 52.34 & 53.86 \\ 
\hdashline 
GMSDI \cite{postolache2024generalized} & 0.37 & 193.54 & 7.73 & 45.29 & 46.67 & 47.50 \\
SA ControlNet  \cite{evans2024stable, zhang2023adding}                   & 0.15 &  170.21 & 2.59 & 55.11 & 55.89 & 57.34 \\
Diff-A-Riff \cite{nistal2024diff} & 0.14 & \textbf{12.16} & 0.90 & \textbf{57.34}  & \textbf{58.00} & \textbf{60.00} \\ 
\hdashline
Ground Truth & - & - & - & 57.70 & 58.66 & 60.75 \\
\midrule 
\textbf{MUSDB18-HQ Test} & & & & & & \\
Random & \textbf{0.087}  & \textbf{6.04} & \textbf{0.49} & 49.49 & 51.27 & 51.32 \\ 
\hdashline
GMSDI \cite{mariani2023multi} & 0.44 &  23.06 & 4.99  & 53.51  & 53.90 & 54.51 \\
SA ControlNet \cite{evans2024stable, zhang2023adding} & 0.12  & 12.14 &  0.66 & \textbf{56.41} & \textbf{57.48} & \textbf{58.82} \\
Diff-A-Riff \cite{nistal2024diff} & 0.20 &  119.41 &  1.68 & 53.79 & 54.75 & 56.08
\\
\hdashline
Ground Truth & - & - & - & 56.47 & 57.55 & 58.94 \\
\bottomrule
\end{tabular}
\end{table*}
In the COLA training procedure, positive pairs are (fully mixed) windows belonging to the same track. In COCOLA (Figure \ref{fig:cocola}), they are submixes belonging to the same window. As such, we allow negative pairs to belong to the same track but in different windows. The $r$ ratio has to be chosen well to avoid strong overlaps between windows on the same track. In that case, we could potentially consider (nearly) coherent submixes as negative pairs.

In addition to the standard mel-filterbank representation of the input (as defined in COLA), we propose a variant where the COCOLA encoder operates on a pair of mel spectrogram features obtained via harmonic-percussive separation (HPS) \cite{Fitzgerald10_HarmPercSep_DAFX}. As shown in Section \ref{sec:results}, this factorized input improves performance and helps us better understand which factor (harmony or rhythm) exhibits greater coherence.

\subsection{COCOLA Score}
\label{subsec:cocola_score}


Equipped with the encoder $f_\theta$, we measure the coherence of accompaniments generated by a conditional model $p_\phi(\mathbf{x} \mid \mathbf{y})$, where $\mathbf{y}$ is the conditioning variable and $\mathbf{x}$ a set of stems or a submix. Given $\mathbf{y}$, the model outputs $\tilde{\mathbf{x}} \sim p_\phi(\mathbf{x} \mid \mathbf{y})$. We embed both $\mathbf{y}$ and $\tilde{\mathbf{x}}$ via $f_\theta$, optionally summing stems, obtaining $\mathbf{h}_{\mathbf{y}} = f_\theta(\mathbf{y})$ and $\mathbf{h}_{\tilde{\mathbf{x}}} = f_\theta(\tilde{\mathbf{x}})$. The \textit{COCOLA Score} is then defined as $\text{sim}(\mathbf{h}_{\mathbf{y}}, \mathbf{h}_{\tilde{\mathbf{x}}})$ (Eq.~\eqref{eq:similarity}), representing the similarity between the embeddings, as illustrated in Figure~\ref{fig:overview}.

\section{Experimental Setup}
\label{sec:setup}

\subsection{Datasets} 
\label{subsec:datasets}
 In our experiments, we use four different stem-separated public datasets to train COCOLA. The datasets are MUSDB18-HQ \cite{rafii2019musdb18}, MoisesDB \cite{pereira2023moisesdb}, Slakh2100 \cite{manilow2019cutting} and CocoChorales \cite{wu2022chamber}. For MoisesDB, which does not have predetermined splits, we set a custom 0.8 (train) / 0.1 (validation) / 0.1 (test) split. For CocoChorales, we use the tiny version, comprising a subset of 4000 tracks.

\begin{figure}[t!]
 \centerline{
 \includegraphics[width=0.8\columnwidth]{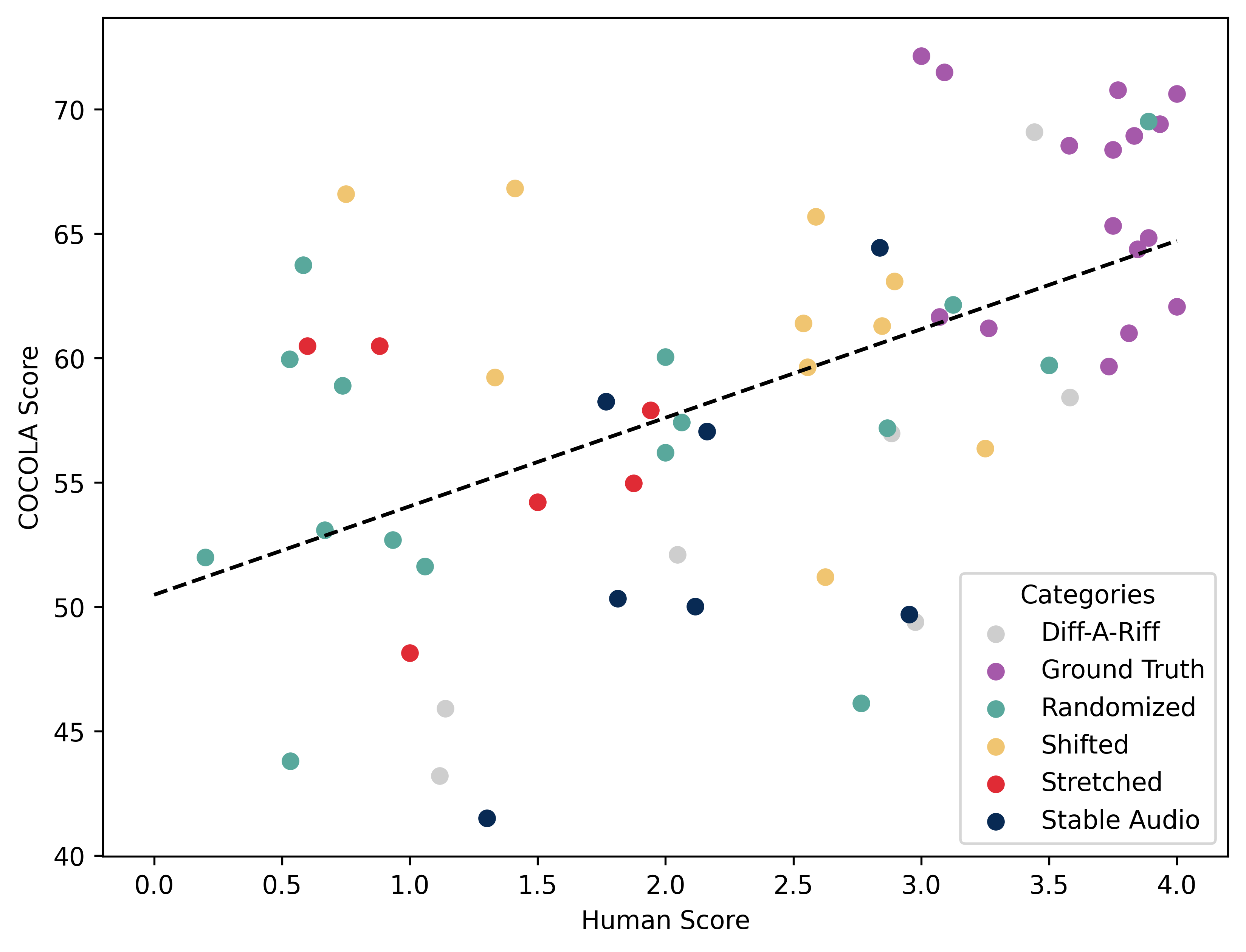}
 }
 \caption{\textbf{Correlation plot} between subjective scores (MOS) ($x$-axis) and COCOLA Scores ($y$-axis).}
 \label{fig:correlation}
\end{figure}

\subsection{Model and Training Details}
\label{subsec:models}

To implement the COCOLA encoder $f_\theta$, we follow the COLA framework \cite{saeed2021contrastive} and employ the EfficientNet-B0 \cite{tan2019efficientnet} (using two input channels for HPS) convolutional architecture followed by a linear projection layer. The embedding dimension is 512. Differently from the original baseline, we add a $0.1$ dropout on the EfficientNet layers.

Each training batch contains 32 5s audio chunks (16kHz). We set the maximum window overlap ratio $r = 50\%$ and train with the Adam optimizer with a $10^{-3}$ learning rate. As a data augmentation method, we add Gaussian noise to positive samples, with $\sigma = 10^{-3}$. During HPS training, masking is applied randomly to one or both of the two channels. This approach allows the model to learn the representations of either component independently. 
 
\section{Experiments}
\label{sec:results}

In our experiments, we trained four COCOLA encoder models without HPS: ``COCOLA MoisesDB'', ``COCOLA Slakh2100'', ``COCOLA CocoChorales'' and ``COCOLA All''. The first three are trained on the homonym datasets, while the last one is trained on all three combined. For the ``COCOLA CocoChorales'' we use all the ensembles, while on ``COCOLA All'' we use only the Random ensemble for a more balanced partitioning with respect to the other datasets. Furthermore, we train a ``COCOLA HPS All'' using HPS with the same dataset as ``COCOLA ALL''. To assess generalization, the smallest dataset, MUSDB18-HQ, is reserved as a held-out test set.

\subsection{Coherent Sub-Mix Classification}
\label{subsec:classification}

We cross-test the performance of COCOLA models trained without HPS by classifying coherent pairs on the test splits of our datasets. More specifically, given an encoder $f_\theta$, we iterate over a test set, collecting a batch of $K$ windows $\mathbf{x}^1, \dots, \mathbf{x}^K$ at each step. Following the steps in Section \ref{subsec:method_contrastive} we compute all similarities $\text{sim}(\mathbf{h}^k_1, \mathbf{h}^j_2)$ for $k, j \in [K]$. We define the accuracy over a batch as:
\begin{equation}
\label{eq:accuracy}
    \frac{1}{K} \sum_{k=1}^{K}\mathds{1}\biggl(k = \argmax_{j \in [K]} \text{sim}(\mathbf{h}^k_1, \mathbf{h}^j_2)\biggr)\,,
\end{equation}
where $\mathds{1}$ is the indicator function. 

We compute the final accuracy by averaging across all batches. Using $K = 2$, Table \ref{tab:classification} shows the results for various combinations of models and test data sets. While "COCOLA MoisesDB" and "COCOLA Slakh2100" perform slightly better than random, "COCOLA Coco-Chorales" shows improved performance. Combining all three datasets achieves over 90\% accuracy.

Adding HPS to ``COCOLA All'', all metrics improve except on CocoChorales (which is already high), reaching 93.87\% on the held-out MUSDB18-HQ and showcasing the usefulness of the factorized representation. We also investigate the performance of ``COCOLA HPS All'' increasing $K$ in Table  \ref{tab:classification_batch}.

\subsection{Accompaniment Generation Evaluation}
\label{subsec:accomp_eval}
For the evaluation of generated music accompaniments, we compare GMSDI \cite{postolache2024generalized}, Diff-A-Riff \cite{nistal2024diff}, and Stable Audio ControlNet (SA ControlNet). The latter is an implementation of a DiT-based \cite{peebles2023scalable} ControlNet \cite{zhang2023adding} for Stable Audio Open \cite{evans2024stable}, which we release\footnote{\url{https://github.com/EmilianPostolache/stable-audio-controlnet}}, given the lack of state-of-the-art open-source models for music accompaniment generation (only MSDM \cite{mariani2023multi} is available, but it is trained on Slakh2100 \cite{manilow2019cutting} not being applicable in realistic settings).

GMSDI is trained on MTG-Jamendo \cite{bogdanov2019mtg}, Diff-A-Riff on a proprietary dataset and we finetune SA ControlNet on MoisesDB. We also consider a Random baseline, where, for a given input, we output a random stem from a different test chunk. 
We evaluate both models on the MoisesDB and MUSDB18-HQ test splits.

We sample 168 chunks ($\sim$47.55s) from the MoisesDB test split and 236 chunks of the same length from the MUSDB18-HQ test split, using random stem subsets as input and querying complementary stems, excluding vocals from the output and the "Other" class in MUSDB18-HQ. SA ControlNet processes the full chunk, while GMSDI and Diff-A-Riff use shorter initial sub-chunks ($\sim$23.78s and $\sim$9.58s, respectively) due to reduced context windows. The outputs are cut to the initial $\sim$9.58s.

We compare the COCOLA Score (based on "COCOLA HPS All") with the FAD \cite{roblek2019fr, gui2024adapting} metric (interpreted as a sub-FAD \cite{donahue2023singsong, mariani2023multi}), computed using CLAP \cite{elizalde2023clap}, EnCodec \cite{defossez2023high}, and VGGish. Results in Table \ref{tab:generative}
 show that FAD tends to favor the Random baseline, as it evaluates perceptual quality but not track coherence, which is better captured by COCOLA. On Ground Truth, we compute COCOLA upper bounds with real positive pairs. Diff-A-Riff performs best on MoisesDB and SA ControlNet on MUSDB18-HQ.

\subsection{MOS vs COCOLA Score}

To evaluate COCOLA's effectiveness, we conducted a subjective test with 41 participants who rated (0-4) the coherence of (pitch) Shifted and (time) Stretched chunks. We also used Random, Ground Truth, Diff-A-Riff, and SA ControlNet tracks from the previous experiment. We average the ratings (MOS) for each chunk. Figure \ref{fig:correlation} shows the correlation plot between MOS and COCOLA Score (using "COCOLA HPS All"). The Pearson correlation coefficient is
$r=$ 0.54 $(p < 0.01)$, having high statistical significance.

\section{Conclusion}
\label{sec:conclusion}
In this paper, we proposed COCOLA, a contrastive encoder to evaluate the coherence between musical stems. Classification experiments and human scores demonstrate the effectiveness of the method. We used COCOLA to benchmark music accompaniment models, proposing a new evaluation metric for the task. We aim to improve COCOLA using additional stem-level \cite{sarkar2022ensembleset} or pre-separated large-scale \cite{bogdanov2019mtg} datasets.

\bibliographystyle{IEEEtran}
\bibliography{cocola}

\begin{thebibliography}{10}
\providecommand{\url}[1]{#1}
\csname url@samestyle\endcsname
\providecommand{\newblock}{\relax}
\providecommand{\bibinfo}[2]{#2}
\providecommand{\BIBentrySTDinterwordspacing}{\spaceskip=0pt\relax}
\providecommand{\BIBentryALTinterwordstretchfactor}{4}
\providecommand{\BIBentryALTinterwordspacing}{\spaceskip=\fontdimen2\font plus
\BIBentryALTinterwordstretchfactor\fontdimen3\font minus \fontdimen4\font\relax}
\providecommand{\BIBforeignlanguage}[2]{{%
\expandafter\ifx\csname l@#1\endcsname\relax
\typeout{** WARNING: IEEEtran.bst: No hyphenation pattern has been}%
\typeout{** loaded for the language `#1'. Using the pattern for}%
\typeout{** the default language instead.}%
\else
\language=\csname l@#1\endcsname
\fi
#2}}
\providecommand{\BIBdecl}{\relax}
\BIBdecl

\bibitem{schneider2023mousai}
F.~Schneider, Z.~Jin, and B.~Sch{\"o}lkopf, ``Moûsai: Text-to-music generation with long-context latent diffusion,'' \emph{arXiv preprint arXiv:2301.11757}, 2023.

\bibitem{garcia2023vampnet}
H.~F. Garcia, P.~Seetharaman, R.~Kumar, and B.~Pardo, ``Vampnet: Music generation via masked acoustic token modeling,'' \emph{arXiv preprint arXiv:2307.04686}, 2023.

\bibitem{evans2024fast}
Z.~Evans, C.~Carr, J.~Taylor, S.~H. Hawley, and J.~Pons, ``Fast timing-conditioned latent audio diffusion,'' \emph{arXiv preprint arXiv:2402.04825}, 2024.

\bibitem{evans2024stable}
Z.~Evans, J.~D. Parker, C.~Carr, Z.~Zukowski, J.~Taylor, and J.~Pons, ``Stable audio open,'' \emph{arXiv preprint arXiv:2407.14358}, 2024.

\bibitem{song2021scorebased}
Y.~Song, J.~Sohl-Dickstein, D.~P. Kingma, A.~Kumar, S.~Ermon, and B.~Poole, ``Score-based generative modeling through stochastic differential equations,'' in \emph{International Conference on Learning Representations}, 2021.

\bibitem{ho2020denoising}
J.~Ho, A.~Jain, and P.~Abbeel, ``Denoising diffusion probabilistic models,'' in \emph{Proceedings of the 34th International Conference on Neural Information Processing Systems}, 2020, pp. 6840--6851.

\bibitem{elizalde2023clap}
B.~Elizalde, S.~Deshmukh, M.~Al~Ismail, and H.~Wang, ``Clap learning audio concepts from natural language supervision,'' in \emph{ICASSP 2023-2023 IEEE International Conference on Acoustics, Speech and Signal Processing (ICASSP)}.\hskip 1em plus 0.5em minus 0.4em\relax IEEE, 2023, pp. 1--5.

\bibitem{raffel2020exploring}
C.~Raffel, N.~Shazeer, A.~Roberts, K.~Lee, S.~Narang, M.~Matena, Y.~Zhou, W.~Li, and P.~J. Liu, ``Exploring the limits of transfer learning with a unified text-to-text transformer,'' \emph{Journal of machine learning research}, vol.~21, no. 140, pp. 1--67, 2020.

\bibitem{mariani2023multi}
G.~Mariani, I.~Tallini, E.~Postolache, M.~Mancusi, L.~Cosmo, and E.~Rodol{\`a}, ``Multi-source diffusion models for simultaneous music generation and separation,'' in \emph{The Twelfth International Conference on Learning Representations}, 2024.

\bibitem{han2023instructme}
B.~Han, J.~Dai, X.~Song, W.~Hao, X.~He, D.~Guo, J.~Chen, Y.~Wang, and Y.~Qian, ``Instructme: An instruction guided music edit and remix framework with latent diffusion models,'' \emph{arXiv preprint arXiv:2308.14360}, 2023.

\bibitem{parker2024stemgen}
J.~D. Parker, J.~Spijkervet, K.~Kosta, F.~Yesiler, B.~Kuznetsov, J.-C. Wang, M.~Avent, J.~Chen, and D.~Le, ``Stemgen: A music generation model that listens,'' in \emph{ICASSP 2024-2024 IEEE International Conference on Acoustics, Speech and Signal Processing (ICASSP)}.\hskip 1em plus 0.5em minus 0.4em\relax IEEE, 2024, pp. 1116--1120.

\bibitem{grachten2020bassnet}
M.~Grachten, S.~Lattner, and E.~Deruty, ``Bassnet: A variational gated autoencoder for conditional generation of bass guitar tracks with learned interactive control,'' \emph{Applied Sciences}, vol.~10, no.~18, 2020.

\bibitem{donahue2023singsong}
C.~Donahue, A.~Caillon, A.~Roberts, E.~Manilow, P.~Esling, A.~Agostinelli, M.~Verzetti, I.~Simon, O.~Pietquin, N.~Zeghidour \emph{et~al.}, ``Singsong: Generating musical accompaniments from singing,'' \emph{arXiv preprint arXiv:2301.12662}, 2023.

\bibitem{roblek2019fr}
D.~Roblek, K.~Kilgour, M.~Sharifi, and M.~Zuluaga, ``Fréchet audio distance: A reference-free metric for evaluating music enhancement algorithms,'' in \emph{Proc. Interspeech}, 2019, pp. 2350--2354.

\bibitem{chopra2005learning}
S.~Chopra, R.~Hadsell, and Y.~LeCun, ``Learning a similarity metric discriminatively, with application to face verification,'' in \emph{2005 IEEE Computer Society Conference on Computer Vision and Pattern Recognition (CVPR'05)}, vol.~1, 2005, pp. 539--546 vol. 1.

\bibitem{oord2018representation}
A.~v.~d. Oord, Y.~Li, and O.~Vinyals, ``Representation learning with contrastive predictive coding,'' \emph{arXiv preprint arXiv:1807.03748}, 2018.

\bibitem{favory2020coala}
X.~Favory, K.~Drossos, T.~Virtanen, and X.~Serra, ``{COALA}: Co-aligned autoencoders for learning semantically enriched audio representations,'' in \emph{ICML 2020 Workshop on Self-supervision in Audio and Speech}, 2020.

\bibitem{chen2020simple}
T.~Chen, S.~Kornblith, M.~Norouzi, and G.~Hinton, ``A simple framework for contrastive learning of visual representations,'' in \emph{International conference on machine learning}.\hskip 1em plus 0.5em minus 0.4em\relax PMLR, 2020, pp. 1597--1607.

\bibitem{manco2022learning}
I.~Manco, E.~Benetos, E.~Quinton, and G.~Fazekas, ``Learning music audio representations via weak language supervision,'' in \emph{ICASSP 2022-2022 IEEE International Conference on Acoustics, Speech and Signal Processing (ICASSP)}.\hskip 1em plus 0.5em minus 0.4em\relax IEEE, 2022, pp. 456--460.

\bibitem{manco2022contrastive}
------, ``Contrastive audio-language learning for music,'' in \emph{Ismir 2022 Hybrid Conference}, 2022.

\bibitem{wu2023large}
Y.~Wu, K.~Chen, T.~Zhang, Y.~Hui, T.~Berg-Kirkpatrick, and S.~Dubnov, ``Large-scale contrastive language-audio pretraining with feature fusion and keyword-to-caption augmentation,'' in \emph{ICASSP 2023-2023 IEEE International Conference on Acoustics, Speech and Signal Processing (ICASSP)}.\hskip 1em plus 0.5em minus 0.4em\relax IEEE, 2023, pp. 1--5.

\bibitem{pascual2019learning}
S.~Pascual, M.~Ravanelli, J.~Serr{\`a}, A.~Bonafonte, and Y.~Bengio, ``Learning problem-agnostic speech representations from multiple self-supervised tasks,'' \emph{Interspeech 2019}, 2019.

\bibitem{huang2022masked}
P.-Y. Huang, H.~Xu, J.~Li, A.~Baevski, M.~Auli, W.~Galuba, F.~Metze, and C.~Feichtenhofer, ``Masked autoencoders that listen,'' \emph{Advances in Neural Information Processing Systems}, vol.~35, pp. 28\,708--28\,720, 2022.

\bibitem{jansen2018unsupervised}
A.~Jansen, M.~Plakal, R.~Pandya, D.~P. Ellis, S.~Hershey, J.~Liu, R.~C. Moore, and R.~A. Saurous, ``Unsupervised learning of semantic audio representations,'' in \emph{2018 IEEE international conference on acoustics, speech and signal processing (ICASSP)}.\hskip 1em plus 0.5em minus 0.4em\relax IEEE, 2018, pp. 126--130.

\bibitem{spijkervet2021contrastive}
J.~Spijkervet, J.~Burgoyne \emph{et~al.}, ``Contrastive learning of musical representations,'' in \emph{Ismir 2021}.\hskip 1em plus 0.5em minus 0.4em\relax ISMIR, 2021.

\bibitem{saeed2021contrastive}
A.~Saeed, D.~Grangier, and N.~Zeghidour, ``Contrastive learning of general-purpose audio representations,'' in \emph{ICASSP 2021-2021 IEEE International Conference on Acoustics, Speech and Signal Processing (ICASSP)}.\hskip 1em plus 0.5em minus 0.4em\relax IEEE, 2021, pp. 3875--3879.

\bibitem{garoufis2023multi}
C.~Garoufis, A.~Zlatintsi, and P.~Maragos, ``Multi-source contrastive learning from musical audio,'' \emph{arXiv preprint arXiv:2302.07077}, 2023.

\bibitem{chen2020neural}
B.-Y. Chen, J.~B. Smith, and Y.-H. Yang, ``Neural loop combiner: Neural network models for assessing the compatibility of loops,'' \emph{arXiv preprint arXiv:2008.02011}, 2020.

\bibitem{huang2021modeling}
J.~Huang, J.-C. Wang, J.~B. Smith, X.~Song, and Y.~Wang, ``Modeling the compatibility of stem tracks to generate music mashups,'' in \emph{Proceedings of the AAAI Conference on Artificial Intelligence}, vol.~35, no.~1, 2021, pp. 187--195.

\bibitem{lattner2022samplematch}
S.~Lattner, ``Samplematch: Drum sample retrieval by musical context,'' in \emph{Ismir 2022 Hybrid Conference}, 2022.

\bibitem{riou2024stem}
A.~Riou, S.~Lattner, G.~Hadjeres, M.~Anslow, and G.~Peeters, ``Stem-jepa: A joint-embedding predictive architecture for musical stem compatibility estimation,'' \emph{arXiv preprint arXiv:2408.02514}, 2024.

\bibitem{nistal2024diff}
J.~Nistal, M.~Pasini, C.~Aouameur, M.~Grachten, and S.~Lattner, ``Diff-a-riff: Musical accompaniment co-creation via latent diffusion models,'' \emph{arXiv preprint arXiv:2406.08384}, 2024.

\bibitem{jayaram2020source}
V.~Jayaram and J.~Thickstun, ``Source separation with deep generative priors,'' in \emph{International Conference on Machine Learning}.\hskip 1em plus 0.5em minus 0.4em\relax PMLR, 2020, pp. 4724--4735.

\bibitem{zhu2022music}
G.~Zhu, J.~Darefsky, F.~Jiang, A.~Selitskiy, and Z.~Duan, ``Music source separation with generative flow,'' \emph{IEEE Signal Processing Letters}, vol.~29, pp. 2288--2292, 2022.

\bibitem{postolache2023latent}
E.~Postolache, G.~Mariani, M.~Mancusi, A.~Santilli, L.~Cosmo, and E.~Rodol{\`a}, ``Latent autoregressive source separation,'' in \emph{Proceedings of the AAAI Conference on Artificial Intelligence}, vol.~37, no.~8, 2023, pp. 9444--9452.

\bibitem{postolache2024generalized}
E.~Postolache, G.~Mariani, L.~Cosmo, E.~Benetos, and E.~Rodolà, ``Generalized multi-source inference for text conditioned music diffusion models,'' in \emph{ICASSP 2024 - 2024 IEEE International Conference on Acoustics, Speech and Signal Processing (ICASSP)}, 2024, pp. 6980--6984.

\bibitem{zhang2024instruct}
Y.~Zhang, Y.~Ikemiya, W.~Choi, N.~Murata, M.~A. Mart{\'\i}nez-Ram{\'\i}rez, L.~Lin, G.~Xia, W.-H. Liao, Y.~Mitsufuji, and S.~Dixon, ``Instruct-musicgen: Unlocking text-to-music editing for music language models via instruction tuning,'' \emph{arXiv preprint arXiv:2405.18386}, 2024.

\bibitem{wang2024audit}
Y.~Wang, Z.~Ju, X.~Tan, L.~He, Z.~Wu, J.~Bian \emph{et~al.}, ``Audit: Audio editing by following instructions with latent diffusion models,'' \emph{Advances in Neural Information Processing Systems}, vol.~36, 2024.

\bibitem{pereira2023moisesdb}
I.~G. Pereira, F.~Araujo, F.~Korzeniowski, and R.~Vogl, ``Moisesdb: A dataset for source separation beyond 4 stems,'' in \emph{Ismir 2023 Hybrid Conference}, 2023.

\bibitem{manilow2019cutting}
E.~Manilow, G.~Wichern, P.~Seetharaman, and J.~Le~Roux, ``Cutting music source separation some {Slakh}: A dataset to study the impact of training data quality and quantity,'' in \emph{Proc. IEEE Workshop on Applications of Signal Processing to Audio and Acoustics (WASPAA)}.\hskip 1em plus 0.5em minus 0.4em\relax IEEE, 2019, pp. 45--49.

\bibitem{wu2022chamber}
Y.~Wu, J.~Gardner, E.~Manilow, I.~Simon, C.~Hawthorne, and J.~Engel, ``The chamber ensemble generator: Limitless high-quality mir data via generative modeling,'' \emph{arXiv preprint arXiv:2209.14458}, 2022.

\bibitem{zhang2023adding}
L.~Zhang, A.~Rao, and M.~Agrawala, ``Adding conditional control to text-to-image diffusion models,'' in \emph{Proceedings of the IEEE/CVF International Conference on Computer Vision}, 2023, pp. 3836--3847.

\bibitem{Fitzgerald10_HarmPercSep_DAFX}
D.~FitzGerald, ``Harmonic/percussive separation using median filtering,'' in \emph{Proceedings of the International Conference on Digital Audio Effects ({DAFx})}, Graz, Austria, 2010, pp. 246--253.

\bibitem{rafii2019musdb18}
Z.~Rafii, A.~Liutkus, F.-R. Stöter, S.~I. Mimilakis, and R.~Bittner, ``Musdb18-hq - an uncompressed version of musdb18,'' Aug. 2019.

\bibitem{tan2019efficientnet}
M.~Tan and Q.~Le, ``Efficientnet: Rethinking model scaling for convolutional neural networks,'' in \emph{International conference on machine learning}.\hskip 1em plus 0.5em minus 0.4em\relax PMLR, 2019, pp. 6105--6114.

\bibitem{peebles2023scalable}
W.~Peebles and S.~Xie, ``Scalable diffusion models with transformers,'' in \emph{Proceedings of the IEEE/CVF International Conference on Computer Vision}, 2023, pp. 4195--4205.

\bibitem{bogdanov2019mtg}
D.~Bogdanov, M.~Won, P.~Tovstogan, A.~Porter, and X.~Serra, ``The mtg-jamendo dataset for automatic music tagging,'' in \emph{Machine Learning for Music Discovery Workshop, International Conference on Machine Learning (ICML 2019)}, Long Beach, CA, United States, 2019.

\bibitem{gui2024adapting}
A.~Gui, H.~Gamper, S.~Braun, and D.~Emmanouilidou, ``Adapting frechet audio distance for generative music evaluation,'' in \emph{ICASSP 2024-2024 IEEE International Conference on Acoustics, Speech and Signal Processing (ICASSP)}.\hskip 1em plus 0.5em minus 0.4em\relax IEEE, 2024, pp. 1331--1335.

\bibitem{defossez2023high}
A.~D{\'e}fossez, J.~Copet, G.~Synnaeve, and Y.~Adi, ``High fidelity neural audio compression,'' \emph{Transactions on Machine Learning Research}, 2023.

\bibitem{sarkar2022ensembleset}
S.~Sarkar, E.~Benetos, and M.~Sandler, ``Ensembleset: a new high quality synthesised dataset for chamber ensemble separation,'' in \emph{Ismir 2022 Hybrid Conference}, 2022.

\end{thebibliography}

\end{document}